\documentclass[a4paper,fleqn,usenatbib]{mnras}

\usepackage{mathptmx}

\usepackage[T1]{fontenc}
\usepackage{ae,aecompl}

\usepackage{graphicx}	
\usepackage{savesym}
\usepackage{gensymb}
\usepackage{amsmath}	
\savesymbol{iint}
\savesymbol{iiint}
\savesymbol{iiiint}
\savesymbol{idotsint}
\usepackage{txfonts}
\restoresymbol{TXF}{iint}
\restoresymbol{TXF}{iint}
\restoresymbol{TXF}{iiint}
\restoresymbol{TXF}{iiiint}
\restoresymbol{TXF}{idotsint}

\usepackage{graphicx}
\usepackage{txfonts}
\usepackage{verbatim}
\usepackage{gensymb}
\usepackage{float}
\usepackage{subcaption}
\usepackage{rotating}
\usepackage{longtable}
\usepackage{verbatim}
\usepackage{changepage}
\usepackage{subcaption}
\usepackage{breqn}
\usepackage{natbib}
\usepackage{amsmath}
\usepackage{pdflscape}
\usepackage{caption}
\usepackage{enumerate}
\usepackage{enumitem}
\usepackage{titlesec}

\usepackage{amsmath}
\usepackage[section]{placeins}
\usepackage{pdflscape}
\usepackage{epstopdf}
\usepackage{amssymb}	

\title{Distribution and spectrophotometric classification of basaltic asteroids}

\author[J. A. Mansour]{
Jad-Alexandru Mansour$^{1,2}$\thanks{E-mail:jadmansour96@gmail.com},
M. Popescu$^{3,4,5}$,
J. de Le\'on $^{3,4}$,
J. Licandro$^{3,4}$
\newauthor
\\
$^{1}$ International Centre for Advanced Training and Research in Physics, Magurele 077125, Ilfov, Romania \\
$^{2}$ Faculty of Physics, Bucharest University, 405 Atomistilor str, Magurele 077125, Ilfov, Romania \\
$^{3}$ Instituto de Astrof\'{\i}sica de Canarias (IAC), C/V\'{\i}a L\'{a}ctea s/n, 38205 La Laguna, Tenerife, Spain\\
$^{4}$ Departamento de Astrof\'{\i}sica, Universidad de La Laguna, 38206 La Laguna, Tenerife, Spain \\
$^{5}$ Astronomical Institute of the Romanian Academy, 5 Cu\c{t}itul de Argint, 040557 Bucharest, Romania\\
}

\date{Accepted XXX. Received YYY; in original form ZZZ}

\pubyear{2019}

\begin{document}
\label{firstpage}
\pagerange{\pageref{firstpage}--\pageref{lastpage}}
\maketitle

\begin{abstract}

We aim to determine the distribution of basaltic asteroids (classified as V-types) based on the spectrophotometric data reported in the MOVIS-C catalogue. A total of 782 asteroids were identified. The observations with all four filters (Y, J, H, Ks), available for 297 of these candidates, allow a reliable comparison with the laboratory data of howardite, eucrite, and diogenite meteorites.

We found that the majority of the basaltic candidates ($\approx$ 95$\%$) are located in the inner main belt, while only 29 ($\approx$ 4$\%$) and 8 ($\approx$ 1$\%$) are located in the middle and outer main belt, respectively. A fraction of $\approx$ 33$\%$ from the V-type candidates is associated with the Vesta family (with respect to AstDyS). We also identified four middle main belt V-type candidates belonging to (15) Eunomia family, and another four low inclination ones corresponding to (135) Hertha.

We report differences between the color indices and albedo distributions of the V-type candidates located in the inner main belt compared to those from the middle and outer main belt. These results support the hypothesis of a different origin for the basaltic asteroids with a semi-major axis beyond 2.5 A.U. Furthermore, lithological differences are present between the vestoids and the inner low inclination basaltic asteroids.

The data allow us to estimate the unbiased distribution of basaltic asteroids across the main asteroid belt. We highlight that at least 80$\%$ of the ejected basaltic material from (4) Vesta is missing or is not yet detected because it is fragmented in sizes smaller than 1 km.
  
\end{abstract}




\section{Introduction}

Basaltic asteroids are fragments of large bodies that went through the process of planetary differentiation \citep[e.g.][]{1993Metic..28..161G, 2002aste.book..183G}. This process is defined as the formation of distinct layers in the interior of a body  due to the differences in materials densities. For objects with metal-silicate composition, the differentiation takes place once the melting temperatures of iron-nickel alloys and of silicate solids are reached. Firstly, the eutectic melting of Fe, Ni-FeS-rich fluids occurs at 950$^{\circ}$C, followed by the silicate melts appearing at 1050 - 1150 $^{\circ}$C while complete melting occurs by 1500$^{\circ}$C  \citep[][and references there in]{2006mess.book..733M,2015aste.book..533S}. If complete melting took place, the body formed is fully differentiated and presents an iron nucleus, a silicate mantle and a basaltic crust. Otherwise, if the surface layer of silicate does not achieve melting temperatures and only the Fe, Ni-FeS interior does, the body is said to be partially differentiated.

For obtaining these temperatures, the primary source of heating in the early Solar System was the decay of short-lived radioisotopes $^{26}$Al and $^{60}$Fe \citep{1993LPI....24..577G, 2003ApJ...588L..41T, 2015aste.book..573S, 2015aste.book..533S}. These isotopes have half-lives of 0.73 and 1.5 Myr indicating that the process of differentiation must have taken place shortly after the accretion of planetesimals (during the first few million years after the formation of calcium aluminum rich inclusions). In this context, the distribution of basaltic asteroids across the main asteroid belt provides traces of the differentiation  process that took place in the early Solar System \citep{2006Natur.439..821B}.

The asteroid (4) Vesta, with a diameter of $\approx$ 500 km, is the largest differentiated asteroid showing a basaltic crust \citep{1970Sci...168.1445M}. It is located at a semi-major axis of 2.36 A.U in the inner main belt and is considered the representative member of basaltic asteroids which have been taxonomically classified as V-type \citep{1984atca.book.....T, 2002Icar..158..146B, 2009A&A...493..283D}. The spectral signature of (4) Vesta  is dominated by two absorption bands in the near-infrared at 1 $\mu$m and 2 $\mu$m. These absorption bands correspond to the presence of olivine and pyroxene mineral groups in which the transition of Fe$^{++}$ ions in the crystalline lattice takes place \citep{1993macf.book.....B}.

Initially, basaltic asteroids have been discovered in the inner main belt and have been linked with (4) Vesta through their similar orbital parameters and surface properties \citep{1993Sci...260..186B}. \cite{1990AJ....100.2030Z} identified the Vesta family of asteroids in the space of proper elements. The objects are called dynamical vestoids. Note that for a basaltic asteroid to be a vestoid it is necessary to fit both dynamical and spectroscopic criteria (i.e to have a spectrum similar to that of Vesta).

An important confirmation of the link between vestoids and Vesta is provided by the discovery of two remnant craters, Rheasilvia  and Veneneia with diameters of 500 $\pm$ 25 and 400 $\pm$ 25 km respectively \citep{1997Sci...277.1492T, 2012Sci...336..694S,2012Sci...336..690M,2012Sci...336..687J}. Results from the Dawn space mission showed that most of the vestan surface has a howardite-eucrite lithology while the largest abundance of diogenitic material can be found in the Rheasilvia region \citep{2013M&PS...48.2166D}. This suggests that the impact managed to strip away a fraction of Vesta's basaltic crust and reveal the upper layers of the mantle \citep{2012Sci...336..697D}.  A fraction of the resulting fragments constitutes the present collisional family while others, through collisions and the Yarkovsky effect, ended up in the regions of orbital resonances. There, their eccentricities were pumped up such that a part of them were ejected from the Solar System or fell into the Sun \citep{1994Natur.371..315F}. Another fraction ended up as near-earth asteroids, and subsequently, part of them are at the origin the howardite-eucrite-diogenite (HED) meteorites.  

A similarity in the spectral signature of (4) Vesta and the HED achondrites lead to the conclusion that Vesta is the parent body of these differentiated meteorites \citep[e.g][]{1977GeCoA..41.1271C, 2002cem..book.....N}. The eucrites make up 52$\%$ of all this group. They are made of anorthite plagioclase (30$\%$-50$\%$) and low Ca pigeonite clinopyroxene (40$\%$-60$\%$), and are associated with the basaltic crust of the differentiated body. Diogenites make up 24$\%$ of HED. Their primary mineral is orthopyroxene and they are associated with the upper mantle layers of the differentiated body. The howardites are a polymict breccias of eucrites and diogenites formed after the impact of a body onto the differentiated asteroid \citep{2002cem..book.....N}.

Recently, basaltic asteroids were further discovered in the middle and outer main belt \citep{2000Sci...288.2033L, 2006astro.ph..9420H, 2009P&SS...57..229D}. Dynamically, it is less probable to link these objects with the Vesta collisional family, due to implausible high ejection velocity needed to transport them from the inner main belt to the outer main belt \citep{2008Icar..193...85N, 2014MNRAS.439.3168C}. The largest of them,  (1459) Magnya \citep{2000Sci...288.2033L}  has an effective diameter of a 17$\pm$1 km and a geometric visible albedo of 0.37$\pm$0.06 \citep{2006Icar..181..618D}. This asteroid is orbiting in the outer main belt at a semi-major axis of 3.15 A.U. The spectral studies showed that its composition deviates, in terms of pyroxene chemistries, from that of (4) Vesta. Thus,  \citet{2004Icar..167..170H} suggested that (1459) Magnya originated from a different parent body, and that its progenitor formed in a more chemically reduced region of the solar nebula within the asteroid belt. 

The discovery of V-types with low orbital inclinations $i$ \citep{2004Icar..171..120D, 2006A&A...459..969A}, which can't be associated dynamically with the Vesta family, shows that the basaltic material is common through the inner Solar System and suggests that other differentiated parent bodies once existed. This idea is supported by the fact that not all the HED meteorites present the same oxygen isotopic compositions \citep{2009LPI....40.2295S}.  

The all sky surveys provide a large amount of data for solar system objects \citep[e.g.][]{2001AJ....122.2749I,2016A&A...591A.115P}. The prominent spectral features of V-type asteroids allowed to identify these bodies even with broad band photometric filters \citep{2006Icar..183..411R, 2017A&A...600A.126L}. They are called V-type candidates (in order to avoid the confusions with the spectrally classified V-types).  The follow-up spectroscopic surveys confirmed them  with a higher probability, in the range of 80 - 90 $\%$ \citep{2010Icar..208..773M,2011A&A...533A..77D, 2014Icar..242..269H, 2015ApJS..221...19H, 2017MNRAS.464.1718M, 2018MNRAS.475..353M, 2018AJ....156...11H, VESTOIDS}. Based on the V-type candidates obtained from the  Sloan Digital Sky Survey (SDSS) spectrophotometric data,  \citet{2008Icar..198...77M} provides a first estimation of the unbiased size-frequency and semi-major axis distribution of basaltic objects. He found that (4) Vesta was the predominant contributor to the basaltic asteroid inventory and he inferred the presence of basaltic fragments in the vicinity of (15) Eunomia. 

Our goal is to determine the distribution of basaltic asteroids based on the near-infrared spectrophotometric data provided by VISTA-VHS survey \citep{2004SPIE.5493..411I,2010ASPC..434...91L, 2012A&A...548A.119C, 2013Msngr.154...35M, 2015A&A...575A..25S} and compiled in the MOVIS-C catalogue. This article continues the work of \cite{2017A&A...600A.126L}. By using the  comparison with RELAB laboratory data we aim to estimate the fractions of howarditic, diogenitic, and eucritic material across the Main Belt. The results provide new constraints for the existence of multiple differentiated primordial bodies.

This work is organized in the following way, in Section 2 we discuss the methods used to analyze the data. These include the selection of the basaltic asteroid candidates, the de-biassing procedure to obtain the absolute magnitude-frequency distribution -- N(H), and the comparison with the laboratory data. In Section 3 we present the statistical results of V-type candidates and the spectrophotometric comparison of these objects with the HED-like compositions. We discuss the N(H) of both basaltic non-vestoids and vestoids  classes and we estimate the total volume and mass of them. The computed values are compared with the results obtained by \citet{2008Icar..198...77M} based on optical SDSS spectrophotometry and with the dynamical determinations of \cite{2008Icar..193...85N, 2015aste.book..297N}. In Section 4, we summarize and discuss the further implications of our results by comparing them with the current scenarios of differentiation.

\section{Methodology}

\subsection{Selection of V-type candidates}

The MOVIS database \citep{2016A&A...591A.115P, 2018A&A...617A..12P} compiles photometric and absolute magnitude data based on VISTA (Visible and Infrared Survey Telescope for Astronomy). This survey covers different areas of the sky by using a set of filters in visible and near infrared (Z, Y, J, H and Ks) in order to study various astrophysical topics like low mass stars, merger history of the Milky Way, properties of Dark Energy and many more. The data processed in the MOVIS catalogues use the VISTA-VHS (VISTA Hemisphere Survey) component of the VISTA that covers the entire southern hemisphere region. This last version of the MOVIS include: 57 Nearth-Earth Asteroids, 431 Mars Crossers, 612 Hungaria asteroids, 51,382 main-belt asteroids, 218 Cybele asteroid, 267 Hilda asteroids, 434 Trojans, 29 Kuiper Belt objects, and Neptune with its four satellites. 

The MOVIS catalogue was partially developed in the framework of Compositional Mapping of Asteroid Population (CMAPS)\footnote{\url{https://observer.astro.ro/cmaps/?description.html}} project. This work uses various data-mining techniques to retrieve spectral and spectrophotometric data, in both NIR and visible regions, from dedicated surveys like VISTA.

\citet{2016A&A...591A.115P} and \citet{2017A&A...600A.126L} have shown that V-type asteroids form a distinct group in the (Y-J) vs (J-Ks) representation. Following their methodology, certain color indices boundaries can be applied in order to select the V-type candidates. However, the catalogues contain objects with photometric errors ranging from milimagnitudes up to the detection limit ($\approx$ 0.4 mag). These errors depend on the asteroid brightness, the atmospheric transparency and the noise of the detectors. The reliability of the selection depends on the cut-off threshold for the colors errors. 

We define the "V-type Main Set" (or VMS) as the selection of the V-type candidates from the MOVIS catalogue obtained by applying the color indices constraints (Y-J) - (Y-J)$_{err} \geq$ 0.45 and (J-Ks) + (J-Ks$_{err}) \leq$ 0.35. These conditions are imposed in accordance with \cite{2017A&A...600A.126L}, who used (Y-J) $\geq$ 0.5 and (J-Ks)$ \leq$  0.3, while we added an error factor as trade off. Furthermore, to increase the  accuracy of the data set we applied the following uncertainty constraints: (Y-J)$_{err} \leq$ 0.15 , (J-Ks)$_{err} \leq $0.15. Finally, the resulting set contains 782 V-type candidates (Table \ref{Candid}). For comparison, \cite{2018A&A...617A..12P} performed a taxonomic classification in which they classified 708 as V-type candidates, 71 objects were unassigned to a taxonomic class and 3 were found as S-type asteroids. These differences in classification occur due to objects having color values close to the transition region between S-types and V-types.

\begin{table*}
\caption{Comparison between the colors classification and other spectral results retrieved from the literature.  The spectral range is shown (Range column). The $Tax_{spec}$ represents the taxonomic class reported in the literature based on the spectra.  The $ HED_{colors} $ represents the results of the comparison with the HED meteorites data based on our NIR colors. The $ HED_{spec} $ represent the category assigned in the literature based on their spectrum. The orbit shows the dynamical category corresponding to the asteroid proper elements. The following references were used, (1) - \citet{2011A&A...533A..77D}, (2) - \citet{2006A&A...459..969A}, (3) - \citet{VESTOIDS}, (4) - \citet{2002Icar..158..146B}, (5) - \citet{2010Icar..208..773M}, (6) - \citet{2016MNRAS.455.2871I}, and (7) - \citet{2017MNRAS.464.1718M}}
\centering
\begin{tabular}{l l c c c l l}
\hline \hline
Designation           & Range & $Tax_{spec}$ & $ HED_{colors} $ & $ HED_{spec} $ & Orbit & Ref. \\
\hline
(2011) Veteraniya     &NIR  & V        & - & H-E & Vestoid & (1) \\
(2486) Metsahovi      &Optic& V        & H & -   & Inner-Other & (2) \\
(2452) Lyot           &VNIR  & V        & D & D   & Middle-Outer & (3)\\
(2508) Alupka         &Optic& V        & H & -   & Vestoid & (4) \\
(2763) Jeans          &VNIR  & V        & E & E   & Low inclination & (1), (5) \\
(3536) Schleiche      &Optic& V        & E & -   & Inner-Other & (4) \\
(3613) Kunlun         &NIR  & V        & - & -   & Vestoid & (6)\\
(3882) Johncox        &VNIR  & V        & E & E   & Low inclination & (3) \\
(3900) Knezevic       &Optic& V        & - & -   & Inner-Other &  (4)\\
(4311) Zguridi        &Optic& V        & H & -   & Vestoid & (4)\\
(4993) Cossard        &NIR  & V        & - & H   & Vestoid & (1) \\
(5051) Ralph          &Optic& Sr/R      & E & -  & Vestoid & (4)\\
(6046) 1991 RF14      &VNIR  & V        & D & H-D & Fugitive & (3)\\
(7459) Gilbertofranco &VNIR & V      & - & E   & Middle-Outer & (3)\\
(9147) Kourakuen      &NIR  & V        & D & D   & Fugitive& (7)\\
(10614) 1997 UH1      &NIR  & V        & E & H-E & Vestoid &  (1)\\
(19281) 1996 AP3      &VNIR  & V        & D & H   & Fugitive& (3)\\
(27343) Deannashea    &VNIR  & V        & D & H   & Low inclination & (5)\\
(40521) (1999 RL95)   &VNIR  & V        & E & -   & Middle-Outer & (5)\\
\hline
\end{tabular}
\label{AsterSpectra}
\end{table*}

The properties of the basaltic candidates are discussed considering their proper orbital elements reported by the AstDyS database \footnote{https://newton.spacedys.com/astdys2/} \citep{2014Icar..239...46M}, and by \cite{2015aste.book..297N} via  Planetary Data System\footnote{https://pds.nasa.gov/}. We used as reference the AstDyS website (accessed on August 5, 2019), which is more conservative in identifying dynamical vestoids.

To improve the characterization of our objects, we searched for other published data. These include albedo, SDSS optical colors, and the few available spectra both in visible and near-infrared.

Thus, we retrieved the asteroids albedos recorded by the WISE survey  \citep{2011ApJ...741...68M, 2014ApJ...792...30M}. This was the case for 330 objects observed in common. Their average value is $\bar p_{V}$ = 0.356 $\pm$ 0.114. This is in accordance with \cite{2011ApJ...741...90M} who found an average value  $\bar p_{V}$ = 0.362 $\pm$ 0.1, while the albedo of Vesta is $p_V$ = 0.36.  Eight ($2.4\%$) out of these V-type candidates have values $p_V \leq$ 0.15, comparable with carbonaceous chondrites compositions, thus it makes them improbable of having a basaltic surface. As a consequence, the low albedo objects were removed from the VMS set.

\citet{2006Icar..183..411R} used the SDSS data to identify the V-type candidates based on the filters in the optical range (u, g, r, i and z). The spectra of V-type asteroids in this range present a steep slope close to 0.7 $\mu m$ and a deep absorption band past 0.75 $\mu m$. In their work, the authors used Principal Component Analysis of the asteroid colors combined with criteria of segregation of the taxonomic classes. We found 36 of their optical V-types candidates in MOVIS-C catalog (the errors were considered within the above mentioned thresholds). A number of 31 ($86\%$) of them are selected by us as basaltic candidates. We mention that the rest of 5 objects in common have (Y-J) and (J-Ks) values at the border of V-types and S-complex.

A more extensive study of the SDSS  database was performed by \citep{2010A&A...510A..43C}. They studied the taxonomic classification and orbital distribution of main belt asteroids based on their colors. By defining nine classes (V, O, Q, S, A, L, D, X and C) linked with the current mineralogical interpretation, they obtained a prediction for each asteroid. We compare our 782 V-types candidates with their data and we find 202 ($\approx 25 \%$ of our data set) common objects. Out of these, 127 have been classified as V-type based on optical SDSS colors, 42 as SQ/ SV/ S and 33 as QV/Q.

Furthermore, we compare our spectrophotometric results with the spectra reported in literature (Table \ref{AsterSpectra}). We found that there are spectra for 19 of our V-type candidates, and 18 ($95\%$) of them are spectrally classified as V-types. The exceptions, (5051) Ralph was classified based on its optical spectrum as Sr type by \citet{2002Icar..158..146B} and reclassified as R type by \citet{2016A&A...591A.115P}.

In order to obtain a reliable matching with the laboratory data, from the total of 782 basaltic candidates we selected those that were observed with all four filters (Y, J, H, Ks). This "V-type Restricted Set" (or VRS) contains a subsample of 297 asteroids. The constraint (H-Ks)$_{err}$ $\leq$ 0.15 was additionally considered. By analyzing the various color-color plots we found that (J-H) vs (H-Ks) plot allows to distinguish the different basaltic groups.

\begin{figure}
\centering
\includegraphics[width=8cm]{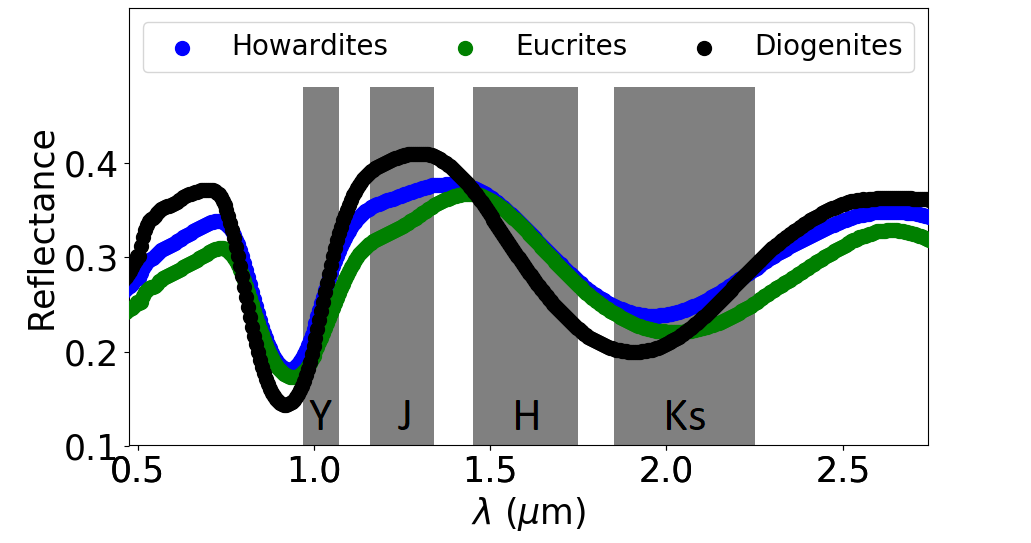}
\caption{The average spectra of the three types of meteorites plotted together with the VISTA filters.}
\label{AverageHEDSpectra}
\end{figure}

\subsection{Comparison with laboratory data}

For the comparison with laboratory data we selected the spectra of howardites, eucrites and diogenites  meteorites from the RELAB database\footnote{\url{http://www.planetary.brown.edu/relab/}} \citep{2004LPI....35.1720P}. This includes a total number of 243 HED meteorites spectra, corresponding to 164 samples out of which 42 (30 samples) are of howardites, 160 (104 samples) of eucrites, and 41 (30 samples) of diogenites. The average spectrum for each type of meteorite is plotted together with the VISTA filters in Fig.~\ref{AverageHEDSpectra}.

The first step for performing the matching between the telescopic observations and the laboratory data, is to convert the meteorites spectra to the VISTA filter system (Y, J, H, Ks). We follow a similar procedure as \cite{2018A&A...617A..12P} briefly summarized it here.

First, the filter transfer function (H$_{F}$) is interpolated at the same wavelengths as the reflectance spectra of meteorites. The product between the filter transfer function and the reflectance values is then integrated in order to obtain the synthetic reflectance corresponding to each band (Eq.~\ref{Reflectconv}). The conversion requires the addition of the colors of the Sun (Eq.~ \ref{Suncolors}). This operation is required because the RELAB data are reflectance spectra while the MOVIS-C observations in this wavelength region correspond to the Sunlight reflected by each asteroid. We used the values (Y-J)$_V$= 0.219 , (J-H)$_V =$ 0.262 , (J-Ks)$_V =$ 0.340 , and (H-Ks)$_V =$ 0.079. These were determined by \citet{ 2018A&A...617A..12P}. They performed the average of the VISTA observations for G2V stars identified based on Two Microns All Sky Survey \citep{2012ApJ...761...16C}

\begin{equation}
R_F =  \int H_{F} (\lambda)* S_{meteor}(\lambda) d\lambda\ \,,
\label{Reflectconv}
\end{equation}

\begin{equation}\label{Suncolors}
C_{FF'} = -2.5\log(R_{F} / R_{F'}) + C_{Sun_{FF'}}\ \,,
\end{equation}
where  $R_F$, $ F \in \left\{Y, J, H, Ks \right\}$ is the reflectance corresponding to filter F, $H_F$ is the filter transfer function, $S_{meteor}$ is the corresponding meteorite reflectance spectrum and $C_{Sun_{FF'}}$ is the Sun color.

To determine the most likely achondrite type (howardite, diogenite, eucrite), we used the K-Nearest Neighbor (KNN) classifier in the (J-H) vs (H-Ks) space. This is an algorithm which attributes a classification label to each new sample by computing the Euclidean distance (used in our case) to $k$ objects with known classification (called training values). The test value will be given the label "x" if it resides in the near proximity (shortest distance) of the training values with that particular label "x". 

\begin{figure*}
\centering
\includegraphics[width=9cm]{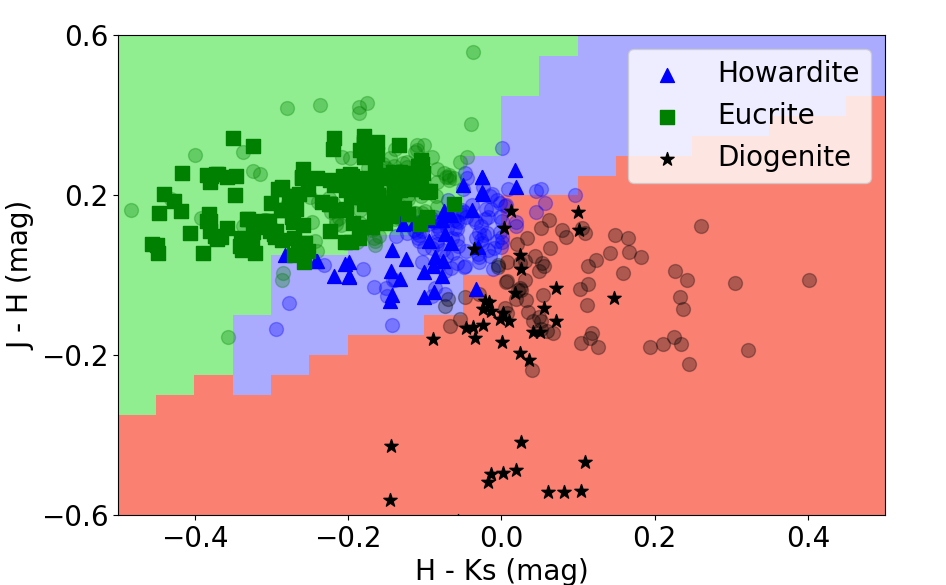}
\includegraphics[width=7cm]{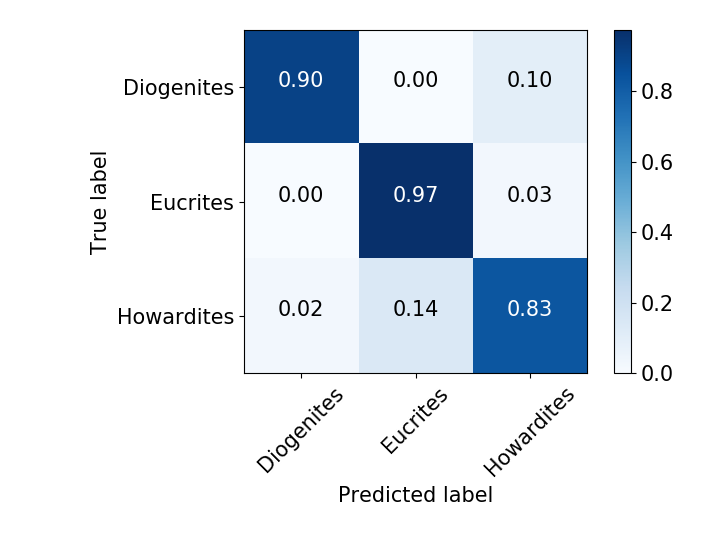}
\caption{Left: The colors of the asteroids (faded, circles) are matched with the colors of the HED meteorites. Right: The confusion matrix performed for the HED data set. The accuracy is indicated by the color gradient.}
\label{AsterandHED}
\end{figure*}

The synthetic colors of the HED meteorites obtained represent the training values for the classification algorithm while the H - howardites, E - eucrites and D - diogenites are the labels. The test values are given by the color indices of the V-type candidates. The algorithm was implemented using \emph{scikit} learn package \cite{scikit-learn}. We merged the results obtained for K = 3, 5, 9 neighbors to assign a label to each V-type candidate. The corresponding regions in the (J-H) vs (H-Ks) were selected since they maximize the separation between H, D and E groups (Fig. \ref{AsterandHED} -- left). 

To test the accuracy of the algorithm we plotted the confusion matrix (Fig. \ref{AsterandHED} -- right) by performing the \emph{Leave One Out Test}. This shows that the algorithm manages to identify with 83$\%$ cases howardites, 97$\%$ eucrites and 90 $\%$ diogenites. The lower percentage of identified howardites is due to their mixed composition of eucrites and diogenites.

In order to account for the uncertainty introduced by the photometric errors we performed a Monte Carlo simulation. For each color we generate 10$^6$ clones based on a Gaussian distribution with the mean represented by the color value and the standard deviation represented by the error. By counting the labels assigned for all cloned colors corresponding to an object we determine the probability of its classification. The results are reported in the online tables.

The classification as H-E-D can be compared with the spectral determinations reported in the literature (Table \ref{AsterSpectra}). For eight objects we have H-E-D determination both based on colors, $ HED_{colors} $ and the spectra $HED_{spec}$. The asteroids (2452) Lyot, and (9147) Kourakuen were found  with a diogenite like composition, both by our work and by \citet{VESTOIDS} and \citet{2017MNRAS.464.1718M}.  We found that (6046) 1991 RF14, (19281) 1996 AP3, and (27343) Deannashea have, according to color classification diogenitic lithology, while the spectral parameters determined by \citet{VESTOIDS} and \citet{2010Icar..208..773M} plot them at the border of H - D region (according to the band I center vs band II center plot defined by \citet{2010Icar..208..773M}). The asteroids (2763) Jeans and (3882) Johncox are eucritic like according to our work and to the spectral data of \citet{2010Icar..208..773M,2011A&A...533A..77D} while (10614) 1997 UH1 is placed in the H - E category. This comparison between our data and the spectral determinations outlines the fact that there is no misidentification between the eucrites and diogenites. The misidentifications with the howardites are explainable for the objects with values close to border regions, considering the errors, or because of the border definition (as howardites are mixtures of eucrites and diogenites).

\subsection{Absolute magnitude frequency distribution of basaltic asteroids}

In this section we present the procedure used to compute the unbiased absolute magnitude frequency distribution of basaltic asteroids, based on the MOVIS-C catalogue. The results are shown in Section 3 where they are compared with the ones obtained by \citet{2008Icar..198...77M} based on the SDSS set, and with the dynamical vestoids defined by \citet{2015aste.book..297N}. We follow a methodology similar with the one described by \citet{2002aste.book...71J} and subsequently by \citep{2008Icar..198...77M}.

\begin{figure}
\centering
\includegraphics[width=9cm]{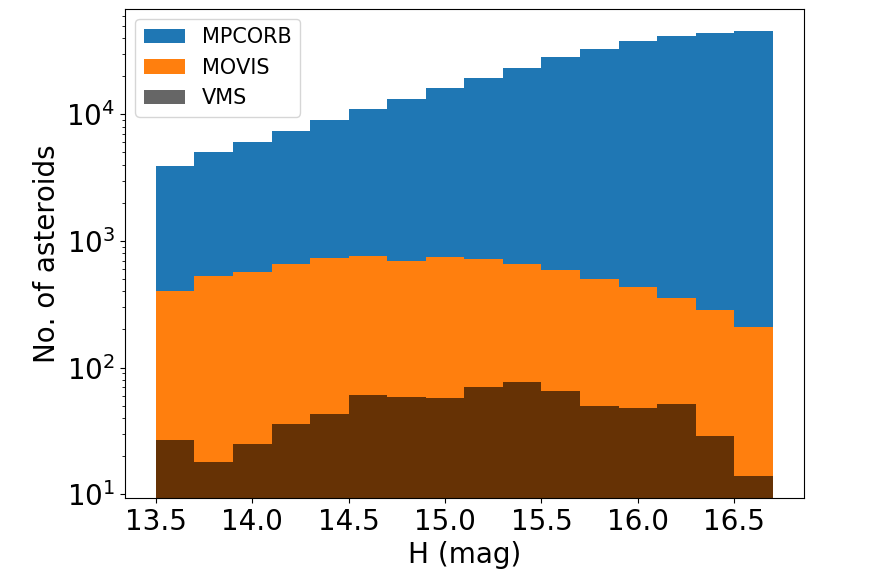}
\caption{Absolute magnitude distribution of V-type candidates with respect to VISTA and MPCORB data base.}
\label{histoall}
\end{figure}
 
In order to determine the absolute magnitude frequency distribution N(\textit{H}), where \textit{H} is the absolute magnitude, we plot in Figure \ref{histoall} the histograms corresponding to the following three data sets: 1) all the asteroids reported by Minor Planet Center Orbit (MPCORB, version November 16, 2018) database \footnote{\url{https://minorplanetcenter.net/}}, 2) the MOVIS-C data with the errors smaller than our selection thresholds,  and 3) the V-type candidates obtained here as the VMS set.

The VMS set of 782 basaltic candidates spans an absolute magnitude interval between 11.4 and 18. We confined this interval between 13.6 and 16.6 in order to avoid the skewness of the distribution that can appear due to the low number of asteroids at the edges of the interval. We use a bin of 0.2 magnitude width as a trade-off between the number of points and the asteroids corresponding to each bin (Fig.~\ref{histoall}). For each bin we obtain a fraction of the V-type candidates relative to the asteroids reported in the MOVIS catalogue. This fraction is then multiplied by the total number of asteroids discovered (the MPCORB set) as it is shown in Eq.~\ref{Nunb}

\begin{equation}
N_{unbiased}^i = \frac{N_{VMS}^i}{N_{MOVIS}^i} * N_{MPCORB}^i\ \,,
\label{Nunb}
\end{equation}
where i represents the bin number (in this case, i = 1, 2, ...n bins), N$_{unbiased}^i$ is the unbiased number of asteroids, N$_{VMS}^i$ is the number of V-type candidates in our set, N$_{MOVIS}^i$ is the number of asteroids in the MOVIS-C data and  N$_{MPCORB}^i$ is the number of asteroids in the MPCORB database (corresponding to each bin). 

The results we obtained follow an exponential law. Thus, we can apply the methodology of \citep{2002aste.book...71J} in order to obtain the total unbiased number distribution of V-type asteroids by fitting with the function $N(H) = k10^{\alpha H}$, where $\alpha$ is the slope while k is a constant. The total unbiased number (Eq.~\ref{Nunbint}) is obtained by integrating over the interval of interest

\begin{equation}
N_{unbiased} = \int k*10^{\alpha H} dH\ \,.
\label{Nunbint}
\end{equation}

In order to compute the volume of the basaltic asteroids, we make use of the empiric formula (Eq.~\ref{Diam}) that relates the diameter ($D$) of an asteroid to its albedo ($p_V$) and absolute magnitude, $H$ \citep{1989aste.conf..524B}

\begin{equation}
D = \frac{1347 km}{\sqrt{p_V}} * 10^{-0.2H}\ \,.
\label{Diam}
\end{equation}

\begin{table*}
\caption{Dynamical categories of the V-type candidates together with the mean value of color indices and albedo. The total numbers of asteroids for which various observations are available are shown.}
\centering
\begin{tabular}{l l l l l l l l l}
\hline \hline
Category         & N_{YJK} & N_{YJHK} & Y-J   				& J-Ks  			& H-Ks 				& J-H 			  & N_p$_V$ & p$_V$  \\ \hline
Vestoids         & 263 		& 102  		& 0.62 $\pm$ 0.07  	& 0.06 $\pm$ 0.12 	&-0.07 $\pm$ 0.14	&0.12 $\pm$ 0.11  &110	&0.37 $\pm$ 0.05  \\ 
Fugitives        & 142  	& 50   	& 0.66 $\pm$ 0.10 	& 0.02 $\pm$ 0.13 	& -0.08 $\pm$ 0.13	&0.12 $\pm$ 0.15  &44 	&0.37 $\pm$ 0.05 \\ 
Low inclination  & 104 	& 40  		& 0.69 $\pm$ 0.12 	& 0.01 $\pm$ 0.12	&-0.07 $\pm$ 0.15 	&0.07 $\pm$ 0.14  &51 	&0.35 $\pm$ 0.06  \\ 
IO    			 & 236   	& 94  		& 0.67 $\pm$  0.10 	& 0.02 $\pm$ 0.11	&-0.06 $\pm$ 0.11	&0.09 $\pm1$ 0.12  &96	&0.36 $\pm$ 0.05 \\ 
MMB $\&$ OMB     & 37 	    & 11        & 0.57 $\pm$ 0.09 	& 0.17 $\pm$ 0.09 	& 0.04  $\pm$ 0.13 	&0.18 $\pm$ 0.16  &21	&0.27 $\pm$ 0.05\\ \hline
\end{tabular}
\label{Dyncat}
\end{table*}

If we assume an albedo of $p_V$ = 0.36 and a spherical body, we can rewrite Eq.~\ref{Diam} in terms of the volume as Eq.~\ref{Mass}
\begin{equation}
V(H) = (1.28 * 10^{18} kg) \frac{10^{-0.6H}}{p_V^{3/2}}\ \,.
\label{Mass}
\end{equation}

\section{Results}

\begin{figure}
\centerline{\includegraphics[width=\hsize]{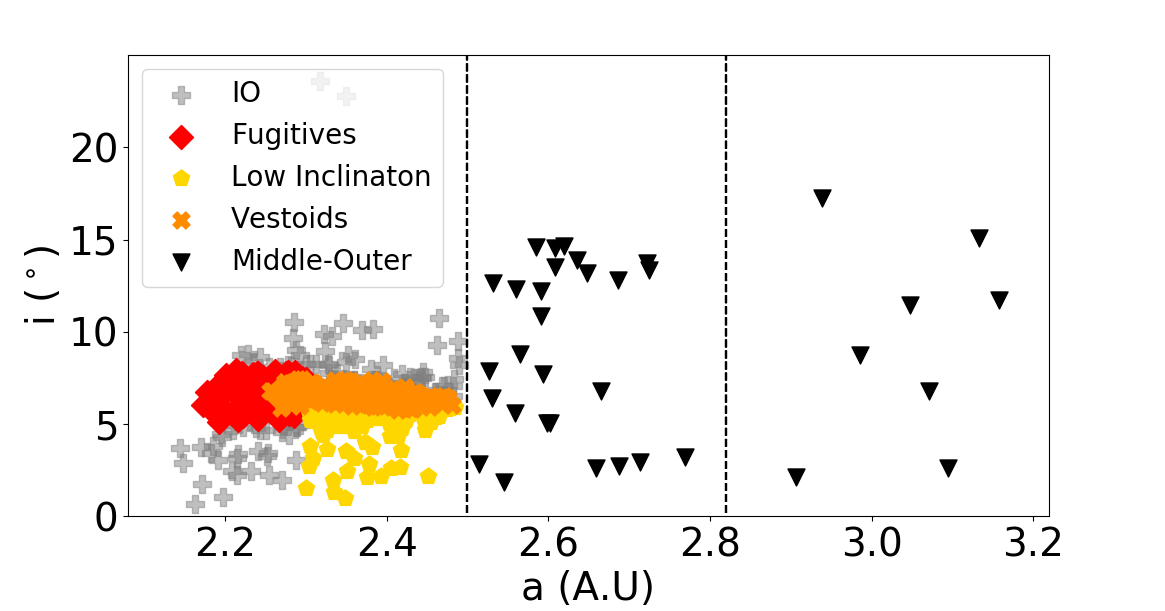}}
\caption{The orbital distribution of the V-type candidates (VMS set) with respect to the five dynamical categories,  plotted with different colors. The location of the most representative resonances with Jupiter are shown.}
\label{MMB_classes}
\end{figure}

This section describes the results with respect to different dynamical categories. The total number of basaltic candidates found in MOVIS-C catalogue are summarized in Table~\ref{Dyncat}. Fig.~\ref{MMB_classes} shows their distribution in the proper elements space of semi-major axis and inclination. The estimated number of basaltic asteroids and their absolute magnitude frequency distribution is determined for the range of absolute magnitudes from 12.1 to 18.3. The classification as H-E-D, of the basaltic candidates allows to infer the fraction of objects corresponding to each of these compositions. This is relevant for tracing their origin to the specific layer of the crust of differentiated parent bodies. 

AstDyS database contains a number of 10\,612 dynamical vestoids with absolute magnitudes between 3.2 and 19.3. Our sample of basaltic candidates contains 263 ($33\%$ from the VMS set) of these bodies. The remaining 519 asteroids are categorized as non-vestoid basaltic asteroids and include \citep{2008Icar..193...85N}: i) the fugitives, which are V-type asteroids with \textit{a} < 2.3 A.U. (where \textit{a} is the semi-major axis) and similar \textit{e} (eccentricity) and \textit{i} (inclination) as the Vesta family; ii) the low inclination V-types are asteroids having \textit{i} $\leq$ 6$\degree$ and 2.3 $<$ \textit{a} $<$ 2.5 A.U. iii) The remaining asteroids in the IMB are named inner-other (IO). The MMB are asteroids with 2.5 < \textit{a} < 2.82. The outer main belt (OMB) are asteroids with \textit{a} > 2.82 A.U. The average color indices and the albedos are shown in Table~\ref{Dyncat} to allow the comparison between these groups. For comparison, \citet{2015aste.book..297N} identified a number of 15\,252 dynamical vestoids with absolute magnitudes between 3.2 and 18.3.

\begin{figure*}
\centering
\includegraphics[width=8cm]{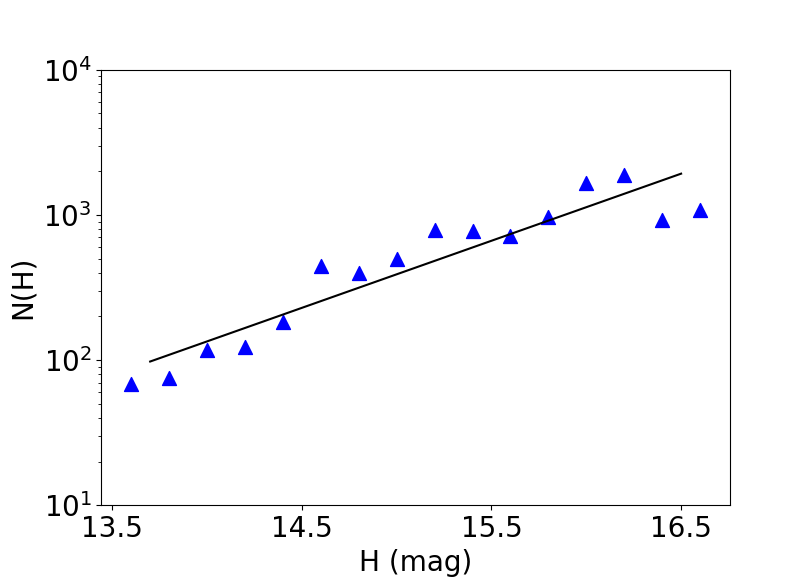}
\includegraphics[width=8cm]{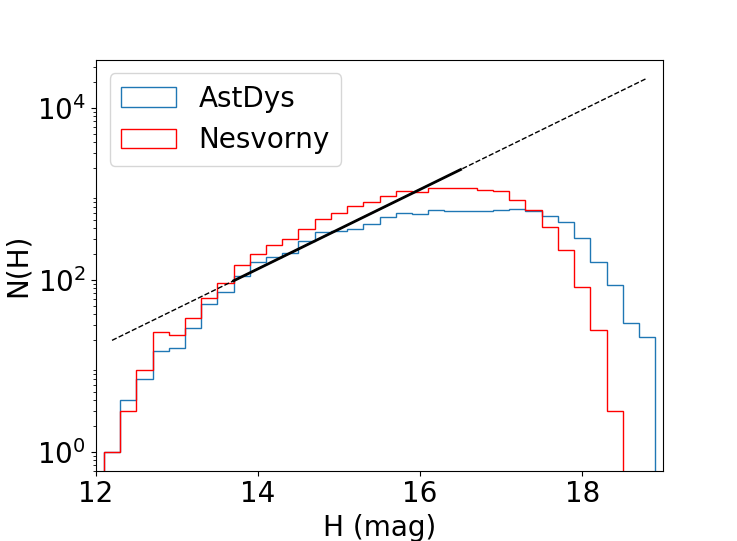}
\caption{Left: The plot shows the unbiased distribution fitted with a power law function N(H) = k10$^{ \alpha H}$. Right: The plot displays a comparison between our unbiassed distribution N(H) -- black (the solid line covers the magnitude interval used for fitting and the dashed line is the extrapolation of the function),the dynamical vestoids from the AstDys database -- blue, and the dynamical vestoids compiled by \citet{2015aste.book..297N}.}
\label{vestodebias}
\end{figure*}

\subsection{The (4) Vesta family of asteroids}

The absolute magnitude range of the (4) Vesta family basaltic candidates covers the interval 12.8 to 17 mag. This corresponds to a diameter range of 0.9 km to 6.1 km (assuming an albedo $p_V$ = 0.36).  The first step in analyzing this dataset is to estimate the unbiased number of vestoids, their total volume and their mass, by following the methodology described in Section 2.3.

In order to estimate the unbiased number of basaltic vestoids ($N(\textit{H})_{Vestoids}$), an adapted form of Eq.~\ref{Nunb} was used. This is computed by obtaining the fraction of basaltic vestoid candidates (instead of all basaltic candidates) relative to all MOVIS-C data and then multiplying it with the total number of all asteroids reported in MPCORB database. Over the interval \textit{H} $\in$(13.6, 16.6), the unbiased number of objects as a function of \textit{H} shows an exponential trend Fig.~\ref{vestodebias} -- left (outside the specified \textit{H} interval, the low number of observed objects preclude the computation). To obtain this histogram, an absolute magnitude bin of 0.2 magnitudes was used as a trade off between the number of points and the number of objects corresponding to each bin. The obtained fit is shown in Fig.~\ref{vestodebias} -- left and Eq.~\ref{NHvestoids} 

\begin{equation}
N(\textit{H})_{Vestoids} = 10^{0.46(\pm 0.03)*H-4.33( \pm 0.49)}\ \,.
\label{NHvestoids}
\end{equation}

The integration of Eq.~\ref{NHvestoids}  allows one to estimate the number of  basaltic vestoids for a given magnitude range. Our power law function $N(H)_{Vestoids}$ (Fig.~\ref{vestodebias} -- right)  is compared with the histogram of dynamical vestoids listed by the AstDys (blue histogram) -- \cite{2014Icar..239...46M}, and by \citet{2015aste.book..297N} -- red histogram. We note the match between our prediction based on the spectrophotometric observations and the dynamical identification. Roughly, this is valid over a magnitude range of $H \in (13, 16)$.  The Fig.~\ref{vestodebias} -- right shows that for magnitudes brighter than $\approx$13 a different slope is required to fit the distribution of dynamical vestoids. For magnitudes fainter than $\approx$16 the low discovery completeness shapes the histogram.

For example, to make a fair comparison, in the interval of absolute magnitudes between 13.6 (the brightest absolute magnitude used for our estimation) and 15 (the approximative turnoff point for the AstDyS histogram) we obtain a number of N$_{Vestoids}$ = 1\,418 $\pm$ 238 (the errors were considered by computing the standard deviation for all the values of the estimated number corresponding to various bin sizes).  Over this absolute magnitude range, the AstDyS lists a number of 1\,571 dynamical vestoids, and \citet{2015aste.book..297N} report a total number of 2\,196 dynamical vestoids. The higher number of dynamical vestoids can be explained by the presence non V-type interlopers. \citet{2017A&A...600A.126L} report that $\sim$85$\%$ of the members of the Vesta dynamical family  are V-type asteroids and only  $\sim2\%$  have carbonaceous like composition, thus are unlikely to be members of the family.

 If we extrapolate the $N(\textit{H})_{Vestoids}$ considering an interval \textit{H} $\in$(12.1, 18.3), which is the common magnitude range between the AstDys, \citet{2015aste.book..297N} and our database,  we predict a N$^{extrap}_{Vestoids}$ = 61\,186 $\pm$ 11\,892 vestoids.  The overestimation introduced by the formula at bright magnitudes has negligible effect on this value. The power law function outlines that  a significant number of small (with an effective diameter within 0.5 - 1.5 km range) basaltic objects  belonging to (4) Vesta family are undiscovered.

 The brightest limit we considered for the absolute magnitude range is motivated by the findings of \citet{2008Icar..198...77M}. They summarized the data from SMASS \citep{1995Icar..115....1X, 2002Icar..158..106B} and $S^3OS^2$ \citep{2004Icar..172..179L} spectroscopic surveys, for characterizing the largest members of the (4) Vesta family. Based on these data, they report that all dynamical vestoids with an absolute magnitudes bellow 12.1, including (63) Ausonia (S-type), (556) Phyllis (S-type) and (1145) Robelmonte (T/D/X -types) are interlopers. In order to obtain an estimate for the volume of basaltic material, we integrate the product between the computed $N(\textit{H})$ and $V(\textit{H})$ functions (Eq.~\ref{VVesto})

\begin{equation}
V_{Vestoids} = \int_{12.1}^{18.3} V(\textit{H})N(\textit{H})d\textit{H}\ \,.
\label{VVesto}
\end{equation}

 The result is $V_{Vestoids} = 7.87\pm0.54~x~10^{4}~km^3$ and corresponds to vestoids with effective diameter in the range of 0.5 - 8 km. For comparison, the AstDyS dynamical vestoids in this range of magnitudes have a volume of $V_{AstDyS}$ = 4.24 x 10$^4$ km$^3$, while those listed by \citet{2015aste.book..297N} found a volume of $V_{Nesvorny}$ = 6.78 x 10$^4$ km$^3$. The differences between our prediction and the total volume of the dynamical vestoids are due to: 1) undiscovered objects at faint magnitudes; 2) interlopers within the (4) Vesta familly; and 3) the power law which overestimates the number of large objects (as shown by Fig.~\ref{vestodebias} -- right). The discrepancy due to point 3) can be inferred by comparing the value of the volume we derive using Eq.~\ref{VVesto} with the one of the dynamical vestoids over the absolute magnitude range were the discoveries are almost complet (i.e. H < 13.6). Thus, by removing the overestimated volume, a realistic prediction can be obtained as $V_{Vestoids} = 6.24\pm0.54~x~10^{4}~km^3$. The total mass of basaltic material in the (4) Vesta familly is M$_{Vestoids}$ = (2.03  $\pm$ 0.17) x 10$^{17}$ kg. This is computed by assuming a density of $\rho$ = 3250 kg$/m^3$.

\begin{figure}
\centering
\includegraphics[width=\hsize]{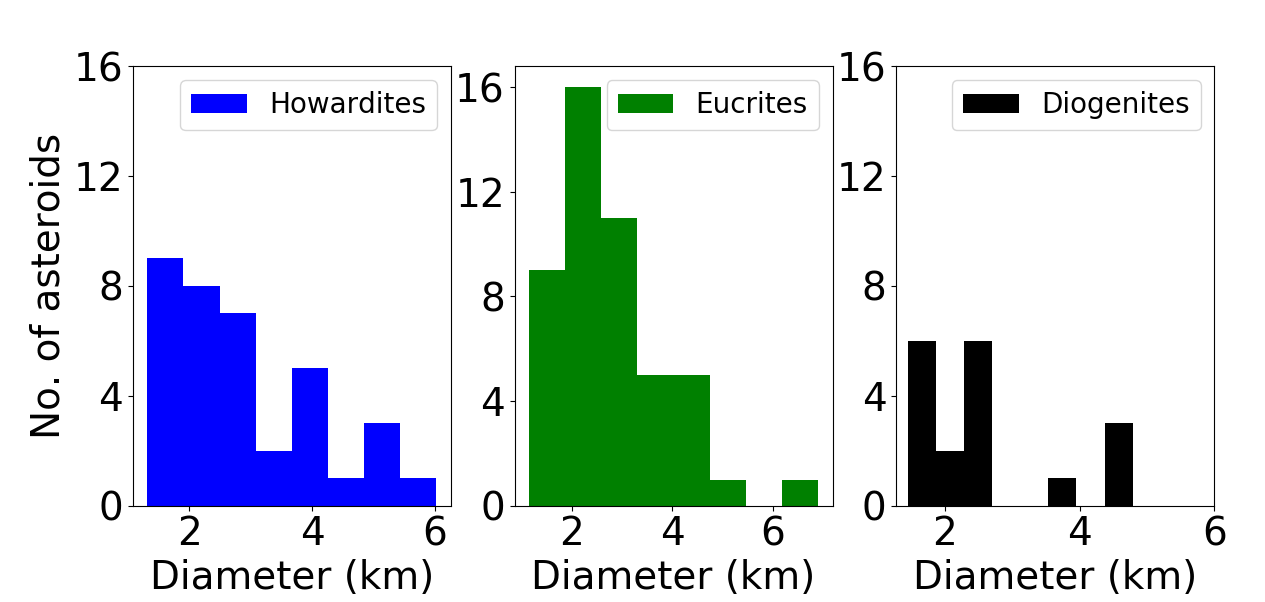}
\caption{The diameter distribution for the vestoid candidates (36 Howardites, 48 Eucrites and 18 Diogenites).}
\label{DistVestoids}
\end{figure}

 \citet{2008Icar..198...77M} performed a similar computation using the data of SDSS survey. One of their objectives was to determine the total mass of the basaltic material in the (4) Vesta family. First, they found a power law of the form $N_{all}(H) = (5.9*10^{-4})*10^{0.5H}$ by fitting the ASTORB distribution between H $\in$ (12.9, 14.8), and a combined efficiency and completeness of the MOC of approximately 22$\%$ up to H = 14.5. This relation is comparable to the one we determined in Eq.~\ref{NHvestoids}, which can be rewritten as $N(H)_{Vestoids} =  (0.47_{-0.15}^{+0.98} *10^{-4}) * 10^{0.46(\pm 0.03)*H}$. They report a total mass of 4.8 x 10$^{16}$ kg, which is four times less than the one we predicted. The differences are explainable by the fact that a different magnitude range was used for determining the power law, and by the numbers of asteroids identified at that time. The inconsistency comes also from the fact that the single slope approximation is not sufficient at bright magnitudes. This is outlined by Fig.~\ref{vestodebias} -- right, and was previously discussed by \citep{2002aste.book...71J}. To chek if our mass estimation is reliable, we  compare it with the total mass of the  dynamical vestoids (with H in the range of 12.1 to 18.3), $M_{AstDyS}$ = 1.38 x 10$^{17}$ kg, $M_{Nesvorny}$ = 2.20 x 10$^{17}$ kg. These values are consistent with the one we found.

There are 102 vestoid basaltic candidates observed with all four filters, thus they can be classified as H-E-D.  Out of these, a number of 48 (47$\%$) are compatible with an eucritic composition, about 36 (35 $\%$) have a howarditic like mineralogy, and 18 (18 $\%$) are compatible with diogenites. The spectral observations of 17 basaltic vestoids performed by \cite{2011A&A...533A..77D} confirmed the predominance of howarditic-eucritic compositions (12 objects), while two diogenitic like asteroids were found.

The ratio of eucrites to diogenites type asteroids together with the size distribution showed in Figure \ref{DistVestoids} is in agreement with \citet{2012LPI....43.2152T, 2013JGRE..118..335M} which suggested that Vesta has a thick basaltic crust of 15-20 km. Our result is also consistent with the spectral analysis in NIR of (4) Vesta obtained by the Dawn mission  which showed that the south-polar Rheasilvia basin display the exposed deeper diogenitic crust compared to the rest of the surface which shows a higher eucritic-howarditic component \citep{2012Sci...336..697D}.

\begin{figure}
\includegraphics[width=\hsize]{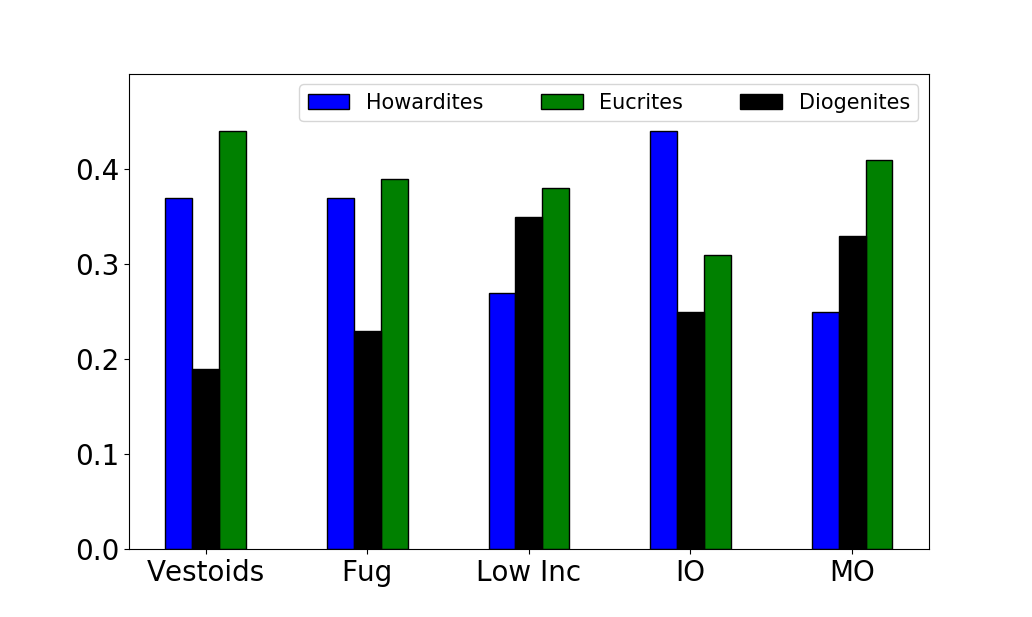}
\caption{ The figure presents the lithological fractions for the  dynamical categories for the basaltic candidates. The color is consistent with the previous figures (blue-Howardite, green-Eucrite and black-Diogenite).}
\label{Fractions}
\end{figure}

\subsection{The non-vestoid basaltic asteroids}

The non-vestoid basaltic asteroids (NVBA) are objects that have been selected as V-type candidates but do not belong to the (4) Vesta family of asteroids in the list of \citet{2015aste.book..297N}. We identify in the MOVIS-C catalog a number of 519 NVBA, and they represent about 66 $\%$ of the total number of basaltic candidates. Out of these objects, 505 ($\sim$ 97 $\%$) are asteroids which are not dynamically linked with any other family of asteroids. There are only 14 bodies associated with other families, including four belonging to (15) Eunomia family, four to (135) Hertha family, and one basaltic candidate in each of the families of (25) Phocaea, (158) Koronis, (170) Maria, (179) Klytaemnestra, (221) Eos, and (2076) Levin.

The large majority of the NVBAs are in the inner main belt, 142 are fugitives, 104 low inclination, and 236 inner-others. There are 37 middle-outer asteroids (Table~\ref{Dyncat}, Fig.~\ref{MMB_classes}).

\begin{figure}
\centering
\includegraphics[width=9cm]{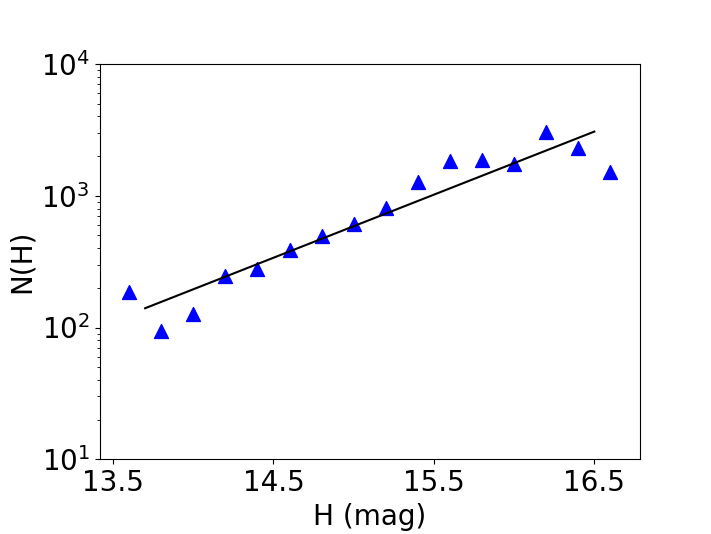}
\caption{Unbiased number of NVBA as a function of magnitude N(H)}
\label{NVBAde}
\end{figure}

In order to estimate the unbiased number of NVBAs as a function of absolute magnitude ($N(\textit{H})_{NVBA}$) we applied the Eq.~\ref{Nunb}. The obtained power function is shown by Eq.~\ref{NVBA} and Fig. \ref{NVBAde} shows the matching

\begin{equation}
N(\textit{H})_{NVBA} = 10^{0.47(\pm 0.03)*\textit{H}-4.41( \pm 0.5)}\ \,.
\label{NVBA}
\end{equation}

 By integrating Eq.~\ref{NVBA} for $H \in (12.1, 18.3)$ we can estimate the expected total number of $N_{NVBA} = 162\,125\pm 27\,868$. The total amount of basaltic material can be approximated in terms of volume and mass (Eq.~\ref{Diam},~\ref{Mass}).  The optimistic extrapolation of Eq.~\ref{NVBA} for \textit{H} $\in$ (12.1 and 18.3) gives a value for the volume of V$_{NVBA}$ = 1.37 $\pm$ 0.1 x 10$^{5}$ km$^3$. Thus, the estimated mass is $M_{NVBA}$ = 4.45 $\pm$ 0.32 x 10$^{17}$ kg (computed for a density $\rho$ = 3250 kg$/m^3$). This result is significantly different compared to \citet{2008Icar..193...85N} because of different absolute interval used for computation, and because of the asteroid discovery completenes at that time compared to now.

A subsample of 195 NVBA has observations with all four filters (Y, J, H, Ks) and can be labeled as H, E, D based on the comparison with the laboratory data. The results show that 74 (39$\%$) are howarditic, 68 (34$\%$) are eucritic and 53 (27$\%$) have diogenitic composition.  For comparison, the spectral studies performed by \cite{2011A&A...533A..77D} for a sample of 24 asteroids found six diogenitic objects, five howarditic objects, two asteroids were labeled as D-H, and 11 as H-E. Their study was focused mostly on the inner main belt asteroids (in their sample they have only one object with semi-major axis higher than 2.5). Both studies highlight that the fraction of diogenitic asteroids is significantly higher for the non-vestoid basaltic bodies. 

Moreover, Fig.~\ref{Fractions} outlines significant differences for the ratio of eucritic to diogenitic asteroids with respect to different dynamical categories. The asteroids with diogenitic like composition represent more than 30$\%$ of the low inclination and middle-outer basaltic candidates compared to (18$\%$) ratio in the basaltic vestoids. This result is an argument for a different origin for the low inclination and middle-outer basaltic candidates. The higher percent of diogenitic material points to violent collisions that disrupted the deeper layers of the differentiated parent body.

{\bf Fugitives}. According to the numerical simulations performed  by \citet{2008Icar..193...85N} a fraction of asteroids can escape from the Vesta family and evolve in their proper $e$ and $i$. These are called "fugitives". Over the time they have dispersed out of the boundaries of the family through various dynamical pathways, including Yarkovsky effect and various mean motion resonances (MMR) like the 1:2 MMR with Mars at 2.42 A.U, 4-2-1 MMR with Jupiter and Saturn at 2.4 A.U, 7:2 at 2.25 A.U and three MMR at 2.3 A.U. We identified 142 basaltic asteroid candidates which can be categorized as fugitives from the Vesta family. We mention that 11 objects have been previously classified by Nesvorny as belonging to the Flora family. A subsample of 50 of them has observations with all four filters, thus 20 (40$\%$) of them are labeled as eucritic, 18 (36$\%$)  as howarditic and 12 (24$\%$) as diogenitic bodies. Compared with other dynamical categories, this HED distribution is the most similar with vestoids. The same conclusion was obtained by \citet{2016MNRAS.455.2871I} which found that the fugitives have similar spectral band parameters with the vestoids.

{\bf Low inclination} \citet{2008Icar..193...85N} proposed that these objects may represent an older family of Vesta, which evolved their inclination during the period of Late Heavy Bombardment epoch ($\approx$ 3.85-3.9 Gyr ago). This is supported by the Ar-Ar radiometric dating of several eucrite meteorites analyzed by \cite{2003M&PS...38..669B} which revealed an age of 4.48 Gyr. A second hypothesis for their origin is that they may be fragments of differentiated bodies other than (4) Vesta \citep{1997M&PS...32..965A, 2008Icar..193...85N}. 

We found 104 basaltic candidates following the definition of low-$i$. A number of 40 of them can be compared with the laboratory data. The result shown that 15 (37$\%$) are compatible with eucritic composition, 14 (35$\%$) with diogenites and 11 (27$\%$) with howardites. This estimation shows that the fraction of diogenitic asteroids with the low $i$ is larger with 17$\%$ than the fraction corresponding to vestoids (Fig.~\ref{Fractions}).

The smallest low inclination basaltic candidate in our sample, (132433) 2002 GZ162, is an eucrite with an equivalent diameter of 1.37 km, while the largest, (2763) Jeans, is also an eucrite with a 7.51 km  diameter (spectral data confirm this finding). Most of the objects we identified in this orbital category have inclinations around 5$^{\circ}$. However, 12 of the basaltic candidates categorized as low-$i$, and nine categorized as inner-others have proper orbital inclinations lower than 3 $^{\circ}$. These are the most difficult to be linked (4) Vesta because they would have required ejection velocity impulse of at least 2 $\frac{km}{s}$, which is implausible \cite{1997M&PS...32..965A}.
 
{\bf Inner-other basaltic asteroids}. This category contains about 30 $\%$ (236) of the total number of basaltic candidates. The four filters observations, available for 94 objects, shows that howarditic compositions (a number of 42 asteroids) are dominating these bodies. About 29 (30$\%$) were found with eucritic composition and 23 (24$\%$) as diogenitic ones (Figure \ref{Fractions}). The smallest IO, (172447) 2003 QX59, is an eucrite with an equivalent diameter of 1.25 km,  while the largest, (2486) Metsahovi, is a howardite with 7.8 km.

{\bf Middle-Outer basaltic asteroids.} In this category are the V-type candidates that can be found in the middle and outer main belt, with a semi-major axis beyond 3:1 resonance with Jupiter. We found 37 basaltic candidates, 29 are located in the middle main belt (MMB), and eight are in the outer main-belt (OMB).  Table~\ref{Dyncat} outlines $\sim1\sigma$ differences between the average colors and albedos of these bodies compared to the inner-main belt ones. The MMB and OMB basaltic candidates show lower albedo, $\bar{p_V}~=~0.27\pm0.05$ compared to $\bar{p_V}~=~0.36\pm0.05$ for inner-main belt V-types, and their colors (Y-J) and (J-Ks) are closer to the S-type border.  These properties suggest a different origin of these bodies than the inner-main belt ones.

\citet{2017MNRAS.468.1236B} determined the probability for asteroids to cross from the inner belt to the middle belt through the 3:1 resonance (at 2.5 A.U)  to be $\approx$ 10$\%$ while the probability of crossing from the middle belt to the outer belt through the 5:2 resonance (at 2.82 A.U) is only $\approx$ 1$\%$. For our data set we find that the ratio of middle to inner belt basaltic candidates is $\approx$ 4$\%$ and of outer to middle basaltic candidates is $\approx$ 27$\%$. These ratios are in contrast with an origin in the Vesta family.

From a total of eleven objects with observations in all four filters, four of them were found as eucritic like, four as diogenitic and three as howarditic. The largest one, (2452) Lyot has an equivalent diameter of 11.9 km, and is located in the outer main belt very close to (1459) Magnya. The comparison with laboratory spectra has labeled it as having diogenitic composition. The spectral studies of \citet{VESTOIDS} confirmed this result and reported that it also shows the lowest wollastonite and forsterite molar contents. They also found that the spectral parameters (band centers and band area ratio) of (2452) Lyot are atypical when compared with other basaltic asteroid spectra.  We also note that (2452) Lyot is the largest basaltic asteroid in our sample.

\section{Discussions and Conclusions}

In this work we studied the distribution of 782 V-type candidates, the largest data set in NIR at the present date, recorded in the VISTA-VHS survey and compiled in the MOVIS-C catalogue.  We found that the vast majority of candidates (about 95$\%$) are located in the inner main belt while the remaining 5$\%$ are spread throughout the middle and outer belt. From a family point of view, more than half (64$\%$) of the V-types have not been associated with other collisional families of asteroids, 33$\%$ have been linked with the Vesta family (the dynamical vestoids) and the rest of 3$\%$ have been associated with other families. These include the inner-main belt family of (8) Flora, (135) Hertha, and the middle-main belt one  of (15) Eunomia. 

We performed a spectrophotometric classification of a subset containing 297 V-type candidates with HED like mineralogies by using a KNN supervised classifier. This classification was performed by correlating the color indices of the candidates with the NIR spectra of a set of 244 HED meteorites selected from the RELAB database. 
 
The results indicate that the vestoids are predominant in eucritic (47$\%$) and howarditic material (35$\%$) followed by a lower fraction of diogenitic like bodies (18$\%$). This is comparable with the findings of the visible and infrared spectrometer on Dawn  spacecraft \citep{2013M&PS...48.2166D} which showed that (4) Vesta has a howardite-rich and eucrite lithology (66.4 $\%$ Howardites / Eucrites, 22.3 $\%$ Eucrites, 7.1 $\%$ Howardites, 4 $\%$ Howardites / Diogenites and 0.2$\%$ Diogenites). 

\citet{1996A&A...316..248M} proposed that the (4) Vesta family of asteroids originated in a collisional event with a body in the range of 40 km diameter. The following observations performed with the Hubble Space Telescope by \citet{1997Sci...277.1492T} supported this scenario. They revealed a crater near the south pole with a diameter of 460 km and a depth of 13 km. The excavation went up to the olivine upper mantle, into the lower crust with high-calcium pyroxene rich composition. They estimated that the ejecta volume was 1.2 x 10$^6$ km$^3$ which is about 1 $\%$ of the total volume of (4) Vesta.

The data obtained by the Dawn spacecraft when visiting (4) Vesta has confirmed this large basin, named Rheasilvia, and evidenced an earlier, underlying crater called Veneneia \citep{2012Sci...336..687J}. The dimensions of the youngest basin, Rheasilvia, were found \citep{2012Sci...336..690M, 2012Sci...336..694S} to be 500 $\pm$ 25 km in diameter and 19 $\pm$ 6 km deep, being slightly larger compared to the ones reported by \citet{1997Sci...277.1492T}. The young crater retention age of this basin indicates that it was formed $\approx$ 1 Gyr ago \citep{2012Sci...336..690M}.  The result is similar with the estimated age for the (4) Vesta family \citep{1997M&PS...32..965A, 2005A&A...441..819C}, thus providing support as being at the origin of the vestoids. The second feature, Veneneia, on the southern pole of (4) Vesta is a semicircular bowl-shaped depression of $\sim$400 $\pm$ 25 km in diameter and a depth of 12 $\pm$ 2 km, half of it destroyed by the Rheasilvia basin. The crater counts suggest an age of 2.1 $\pm$ 0.2 Gyr \citep{2012Sci...336..694S}.

The basaltic candidates belonging to the (4) Vesta family, identified in this work, have diameters bellow 8 km with most of the objects having sizes bellow 5 km. This result is consistent with the size of the Vesta craters. The diogenitic-like asteroids, associated with deeper layers of the crust have the lowest diameters, being smaller than 5 km. 

\citet{2012Sci...336..694S} estimated that the minimum volume of excavated material from Rheasilvia is above $\sim$1 x 10$^6$ km$^3$. Part of it, was retained on the surface and assuming an average ejecta thickness of 5 km over a range of 100 km they roughly estimated the volume of ejecta on the surface as 5 x 10$^5$ km$^3$. Our result is strongly in contrast to these values, the total unbiased amount of basaltic material from vestoid candidates, obtained by extrapolating the Eq.~\ref{VVesto} over the  H $\in$ (12.1, 18.3) interval, is  $V_{Vestoids} = 6.24\pm0.54~x~10^{4}~km^3$.  This value is one order less than the estimated amount of ejecta lost to space from Rheasilvia crater.  By simply extrapolating the power laws down to \textit{H} = 25 (this corresponds to diameter of an asteroid of 22 m size), adds about $\sim$ 25 $\%$ of basaltic material to this value. These results are largely in contrast with the  $\geq$ 5 x 10$^5$ km$^3$ amount of vestan basaltic ejecta, and show that an amount of at least $\approx$4 x 10$^{5}$ km$^3$ basaltic material is missing.

The missing material is an argument for the "battered to bits" scenario proposed by \citet{1996M&PS...31..607B} which states that the fragments of differentiated bodies were continually broken down until their sizes were bellow our observational capabilities. Our analysis is made considering bodies larger than $\sim$ 1 km, thus in order to explain the large amount of missing basaltic material, the N(\textit{H}) function needs to have a very different equation for smaller size asteroids, compared to what we determined. 

One path to investigate the basaltic fragments in the hundred meter size range is the study of near-Earth asteroids (NEAs). \citet{2019Icar..324...41B} showed that the widespread inner belt availability of "Vesta debris" inside the 3:1 resonance with Jupiter gives a high probability to contribute to NEAs population. They reported an overall ratio in the range of $\sim$5$\%$ for the number of V-types NEAs. However,  \citet{2018P&SS..157...82P} and \citet{2019A&A...627A.124P} reported a ratio of $\sim$10$\%$ for the number of V-types NEAs with a size in 100 - 1000 m range. These are comparable with the ratio of V-types in the inner main belt population. 

\citet{2017MNRAS.468.1236B} have performed simulations of the dynamical evolution of V-types. They considered a model of the giant planets migration that took place over the first 700 Myr of the Solar System history. This migration occurred as a result of an interaction between the giant planets and a disk of planetesimals exterior to Neptune's orbit. This model assumes that initially five giant planets existed, Jupiter, Saturn and three ice giants. During the instability caused by the interaction, mutual close encounters of the planets took place causing a Neptune size ice giant to be ejected from the Solar System. As a consequence, asteroid families that formed before or during the migration could have dispersed beyond recognition.

The questions that arise regarding the non-vestoids basaltic candidates are related to their formation. Were they part of the (4) Vesta family in the past and evolved out of the current established orbital boundaries through various dynamical mechanisms? Or maybe they are not related to Vesta at all and are fragments of other bodies that differentiated which are no longer present in the belt.

After the discovery of the basaltic nature of (1459) Magnya, \citet{2002Icar..158..343M} proposed that this asteroid is a fragment of another large differentiated parent body that existed in the outer belt region. The following spectral studies of several authors \citep[e.g.][]{2004Icar..167..170H, 2018AJ....156...11H, 2018MNRAS.479.2607I, VESTOIDS} show different spectral parameters of the basaltic asteroids in the middle and outer main belt supporting the existence of another differentiated parent body. Quantitatively, the number of V-types candidates that we found  to orbit beyond 3:1  resonance with Jupiter is only about $\sim$ 5 $\%$ from the total number of the swarm that surround (4) Vesta (including vestoids, fugitives and IO). Nevertheless, we found 1$\sigma$ difference both in terms of NIR colors and albedo between the vestoids and the basaltic candidates in the middle and outer asteroid belt. In particular, we note the diogenitic nature of the outer belt, 12 km size body, (2452) Lyot, confirmed spectrally by \citet{VESTOIDS}. This is the largest asteroid with a diogenitic nature. 

These differences are in favor of at least another primitive differentiated parent body, which must have been disrupted very early in the history of the Solar System and its fragments dispersed, in order to explain the low number of basaltic asteroids. Moreover, the existence of Lyot, in the outer belt, suggest a more violent collision than the one that produced (4) Vesta family for which we identified diogenitic vestoids only in the order of 5 km size.

\citet{2008Icar..198...77M} identified six basaltic candidate asteroids in the (15) Eunomia family supporting the hypothesis that this parent body is partially or fully differentiated \citep{1997Icar..125..446R, 2005Icar..175..452N}. They suggested that the spectral variations on Eunomia's surface indicate a remnant of a differentiated body that suffered a collisional breakup $\sim$1.3 Gyr forming the differentiated body. Here we report another four basaltic candidates found based on NIR colors. These are  (22032) Mikekoop, (26592) Maryrenfro, (197480) 2005 JW71, (180703) 2004 HW46. All of them have sizes bellow 6 km, and the measured albedo, available for three of them, is in agreement with the V-type composition. \citet{2014MNRAS.439.3168C} has showed through dynamical integration that V-types with low inclination values in the middle main belt could evolve from the parent bodies of the Eunomia and Merxia / Agnia on timescales $\approx$ 2 Gyr.  

The other family on which we found four basaltic candidates is the one of (135) Hertha. These are (143891) 2003 YP43, (176724) 2002 RH2, (102601) 1999 VD5, (434017) 2001 QP329. All of them have diameters in the range of 1-2 km. The asteroid (135) Hertha has a size in the range of 79 km and orbits in inner region of the asteroid belt. It is classified as M type \citep{1984atca.book.....T} or Xk type \citep{2002Icar..158..146B} which suggest that it can be the striped nucleus of a differentiated body. Detailed numerical simulations are required to confirm if the origin of low inclination basaltic candidates is linked with (135) Hertha. Another explanation proposed for the existence of some non-vestoids (low-inclination and fugitives) is that the other major crater on Vesta, Veneneia, could be a potential source \citep{2012Sci...336..694S}.

\section{Acknowledgements}
 The article is based on observations acquired with Visible and Infrared Survey Telescope for Astronomy (VISTA). The observations were obtained as part of the VISTA Hemisphere Survey, ESO Program, 179.A-2010 (PI: McMahon). The work of J.A.M. and part of the work of M.P. was supported by a grant of the Romanian National Authority for Scientific Research - UEFISCDI, project number PN-III-P1-1.2-PCCDI-2017-0371. M.P., J.dL. and J.L. acknowledge support from the AYA2015-67772-R (MINECO, Spain). M.P. and J.dL. also acknowledge financial support from projects SEV-2015-0548 and AYA2017-89090-P (Spanish MINECO).

\bibliographystyle{mnras}
\bibliography{MyBib}

\appendix
\clearpage
\newpage
\onecolumn

\section{Tables of basaltic candidates and their classification}

\begin{table*}
\caption{List of the 782 asteroids basaltic candidates found in this article}
\centering
\begin{tabular}{c c c c c c c c c c c c c}
\hline \hline
1959&9616&17976&27094&35057&44930&53039&64679&75585&96823&123786&166226&403796\\
1979&9746&18012&27239&35062&44953&53050&65027&75636&96890&124440&168519&434017\\
2011&10056&18508&27343&35069&45220&53446&65040&75661&98063&125026&169057\\
2275&10152&18581&27373&35222&45323&53580&65068&77122&98167&125371&170309\\
2452&10613&18644&27383&35284&45327&53590&65392&77244&98297&125390&172447\\
2486&10614&18681&27539&35414&45792&53608&65504&77324&98355&125418&172482\\
2508&11189&18754&27727&35675&45893&53639&65949&77590&98394&126981&172494\\
2763&11326&19025&27770&35792&45942&53703&66111&77972&98454&128565&173551\\
2888&11522&19230&27939&36021&46281&53734&66266&79426&98490&129012&175543\\
3153&11871&19257&28020&36391&46465&53861&67264&79588&98654&130055&175841\\
3188&12172&19281&28217&36475&46527&53870&67416&80351&99579&130718&176159\\
3331&12289&19294&28397&36644&46698&54084&67477&80463&100509&130742&176724\\
3536&12340&19518&28461&36761&47094&54087&67498&80473&101026&130756&177529\\
3613&12591&19573&28543&36834&47206&54215&67748&80525&101029&130797&178112\\
3882&12612&19619&28578&37113&47387&54367&67772&80655&101411&130898&179851\\
3900&12787&19656&28735&37149&47459&54502&67792&80664&101997&130966&179889\\
3954&12789&19679&28737&37192&47476&54667&67850&80798&102571&130972&180441\\
4228&13054&19680&28902&37234&47837&55315&67876&80798&102601&131609&180703\\
4311&13164&19738&29173&37386&48036&55456&68141&80798&102813&131903&180918\\
4444&13191&19931&29186&37404&48112&55700&68318&80805&102986&132347&183004\\
4693&13194&19969&29334&37646&48472&56169&68759&81004&102995&132433&183073\\
4993&13287&19983&29384&37730&48629&56369&68782&82187&103042&132454&186029\\
5051&13380&20071&29450&38127&48644&56456&68801&82241&103105&133202&186771\\
5150&13398&20188&29677&38317&48734&56566&68879&82271&103132&133491&187159\\
5307&13410&20252&29733&38335&48993&56599&69255&82349&103418&133687&188102\\
5328&13530&20254&29834&38403&49119&56653&69742&82455&103848&135575&189231\\
5631&13569&20302&29994&38620&49214&57104&70081&84328&104099&136229&189282\\
5713&13760&21307&30097&38732&49778&57119&70138&84355&106480&136439&191601\\
5758&13855&21633&30176&38744&49884&57233&70145&84357&106640&137211&193561\\
5875&13994&21692&30191&38876&49907&57278&70160&86284&106795&137512&193797\\
5952&14507&21883&30290&38879&49917&57387&70248&86520&107617&137666&196461\\
6014&14809&21936&30329&38906&49950&57454&70277&86560&107709&138192&197480\\
6046&15031&21949&30358&39175&50016&57615&70393&86768&108041&138861&200574\\
6085&15032&22032&30818&39465&50035&57818&70674&87557&108793&139018&200575\\
6096&15121&22080&30893&39476&50048&57930&70694&88438&109080&141517&200690\\
6259&15415&22155&30961&39917&50049&57943&70918&88494&110051&141609&200837\\
6363&15476&22267&31132&39926&50082&58136&70940&88781&111947&141700&201576\\
6406&15506&22409&31460&39949&50084&58271&71010&88958&111976&141740&203379\\
6442&15630&22654&31509&40258&50086&58617&71340&89091&112087&142879&203475\\
6506&15678&22880&31517&40373&50091&58623&71826&89270&112841&143891&203510\\
6581&15706&22892&31575&40521&50105&59215&71850&89271&112981&144220&207473\\
6584&15728&23472&31599&40708&50152&59336&71950&89729&113516&144329&208218\\
6587&15734&23522&31677&41433&50167&59356&71960&90223&113910&145626&210830\\
6853&15781&24050&31775&41463&50215&59423&72027&90442&113914&146305&211140\\
6877&15846&24085&31778&41557&50241&59569&72246&90639&114858&147141&224794\\
7005&15881&24115&32022&41558&50248&59834&72271&90788&116677&147703&224836\\
7012&15885&24140&32276&41776&50472&60394&72304&91343&117819&149666&225334\\
7044&16169&24255&32541&41793&50650&61169&72970&91384&118295&152165&225780\\
7223&16234&24261&33049&41880&51368&61189&73109&92577&118297&152933&227173\\
7459&16274&24538&33100&41896&51411&61241&73170&92581&118305&153392&232170\\
7675&16352&24604&33366&41954&51443&61521&73174&92600&118532&153644&235061\\
7810&16477&24996&33385&42644&51487&61736&73624&92635&119085&153745&238571\\
7823&16605&25220&33477&42656&51511&61741&73658&92646&119136&154716&240551\\
7998&16873&25354&33491&43388&51562&61985&74010&92686&119169&156011&243137\\
8644&17057&25708&33512&43446&51628&61986&74575&93394&119477&156284&244232\\
8790&17139&25979&33513&43885&51687&62061&74596&94355&119969&158281&245176\\
8921&17162&26097&33562&44012&51742&62406&74866&94413&121944&159601&252693\\
9007&17431&26238&33590&44541&52132&62422&74874&94558&122101&161029&283114\\
9064&17546&26401&33628&44569&52819&63353&74898&94680&122243&161033&284907\\
9147&17562&26433&33852&44645&52953&63673&74912&95024&122267&161539&304873\\
9197&17708&26559&33875&44673&52973&63708&74924&95754&122306&162518&322744\\
9204&17739&26592&34081&44682&52985&63997&74936&96653&122392&163726&333396\\
9220&17769&26611&34259&44805&52995&64201&75080&96671&122643&164235&334550\\
9302&17899&26842&34534&44877&53000&64402&75288&96698&123069&165142&354036\\
9495&17934&27081&34650&44878&53037&64458&75289&96750&123381&166121&369457\\
 \hline
\end{tabular}
\label{Candid}
\end{table*}

\begin{table*}
\caption{Classification of basaltic candidates. The designation, the H -E - D classification, and the probability  are shown.}
\centering
\begin{tabular}{|c c c | c c c| c c c| c c c| c c c|}
\hline \hline
|$Des$ & $Cls$ & $Prob$ & $Des$ & $Cls$ & $Prob$ & $Des$ & $Cls$ & $Prob$ & $Des$ & $Cls$ & $Prob$ & $Des$ & $Cls$ & $Prob$ \\ \hline
1979&H&86&50248&H&53&13164&E&90&68141&E&67&42656&D&100\\
2486&H&52&51368&H&89&13191&E&80&68759&E&80&44682&D&100\\
2508&H&100&51742&H&58&13380&E&89&68879&E&100&44953&D&100\\
3331&H&100&52985&H&58&13410&E&98&70160&E&54&45323&D&100\\
3954&H&64&53037&H&46&13855&E&100&70393&E&98&51562&D&69\\
4311&H&72&53446&H&65&15476&E&77&70674&E&84&53590&D&56\\
4444&H&89&53703&H&58&15728&E&84&70694&E&79&53608&D&88\\
4693&H&100&56653&H&74&17057&E&100&71826&E&80&54502&D&92\\
5713&H&81&57233&H&60&17139&E&100&75080&E&100&57278&D&75\\
5758&H&81&57454&H&98&17162&E&63&75636&E&65&59215&D&100\\
6014&H&100&57930&H&51&17708&E&83&82187&E&54&59336&D&100\\
6085&H&93&61741&H&61&18508&E&54&82271&E&76&61189&D&99\\
6096&H&99&61986&H&64&18754&E&58&86284&E&93&62422&D&58\\
6584&H&100&63673&H&51&19025&E&88&86520&E&97&64458&D&53\\
7810&H&93&65027&H&60&19518&E&89&88438&E&91&64679&D&81\\
9197&H&75&67477&H&100&19931&E&92&89270&E&95&67850&D&55\\
9204&H&60&67772&H&56&22032&E&90&89271&E&99&67876&D&99\\
9495&H&99&68801&H&98&22080&E&100&89729&E&95&69255&D&79\\
9616&H&59&70248&H&73&23472&E&86&95754&E&100&70138&D&99\\
12612&H&60&74010&H&92&24115&E&85&98490&E&48&70145&D&99\\
12787&H&86&74898&H&61&24255&E&92&102601&E&100&70918&D&74\\
15506&H&94&79588&H&65&26433&E&100&107709&E&99&71850&D&96\\
15678&H&59&80805&H&49&27094&E&76&130718&E&73&71960&D&93\\
16169&H&66&91384&H&68&27939&E&77&130972&E&96&77972&D&72\\
16477&H&52&98654&H&73&28902&E&60&132433&E&100&84357&D&100\\
18644&H&98&103042&H&58&29334&E&100&165142&E&56&90442&D&53\\
19294&H&75&103132&H&84&30191&E&75&172447&E&53&92600&D&78\\
19738&H&63&111947&H&74&30290&E&87&172494&E&69&93394&D&100\\
19969&H&86&111976&H&66&30358&E&90&193797&E&87&94413&D&90\\
19983&H&100&112841&H&71&31132&E&96&200575&E&47&96653&D&100\\
20302&H&57&112981&H&66&32541&E&74&245176&E&90&96671&D&76\\
21307&H&60&114858&H&67&33477&E&87&2275&D&100&96823&D&67\\
21633&H&56&118532&H&49&33513&E&86&2452&D&100&98394&D&98\\
24085&H&89&122101&H&60&33628&E&100&5952&D&84&101026&D&90\\
24604&H&84&122243&H&69&35284&E&55&6046&D&100&130055&D&96\\
26097&H&56&122267&H&71&36644&E&95&6853&D&100&144329&D&95\\
27727&H&50&136229&H&82&37730&E&72&6877&D&100&168519&D&91\\
27770&H&69&138192&H&55&38127&E&62&7012&D&57\\
28397&H&90&141517&H&51&38879&E&89&7044&D&54\\
29173&H&76&153745&H&58&39476&E&57&9147&D&87\\
29384&H&81&200690&H&55&39917&E&89&13530&D&100\\
29450&H&92&207473&H&66&40521&E&91&15032&D&94\\
29677&H&54&210830&H&69&41558&E&54&15846&D&70\\
30176&H&69&224794&H&53&41954&E&85&16352&D&94\\
30329&H&74&322744&H&58&44541&E&90&17739&D&94\\
30893&H&72&1959&E&74&44645&E&92&17934&D&47\\
31775&H&97&2763&E&95&44877&E&66&18681&D&97\\
31778&H&54&2888&E&100&44878&E&89&19281&D&100\\
33590&H&66&3536&E&100&45893&E&99&19619&D&99\\
34259&H&55&3882&E&95&47459&E&100&20188&D&100\\
34534&H&53&5051&E&100&48112&E&59&20254&D&76\\
35062&H&69&5328&E&90&49119&E&61&22880&D&90\\
36834&H&89&5631&E&100&50086&E&91&22892&D&98\\
38317&H&77&6506&E&95&50091&E&97&27343&D&100\\
38403&H&84&7675&E&100&50152&E&54&27539&D&96\\
41433&H&96&7823&E&100&50650&E&49&30961&D&95\\
41557&H&54&7998&E&100&52132&E&99&31517&D&100\\
42644&H&100&9064&E&85&53000&E&100&31599&D&100\\
43388&H&80&9220&E&79&54367&E&83&33875&D&100\\
44012&H&63&9746&E&100&55315&E&100&34081&D&88\\
44569&H&64&10056&E&97&57104&E&81&36475&D&73\\
44805&H&72&10614&E&93&58271&E&62&37386&D&56\\
47094&H&60&12172&E&94&60394&E&80&39465&D&65\\
48644&H&54&12340&E&85&61736&E&57&39926&D&87\\
50105&H&95&13054&E&100&63708&E&97&41776&D&97\\

 \hline
\end{tabular}
\label{HEDCandid}
\end{table*}

\label{lastpage}
\end{document}